\documentclass{appolb}
\usepackage{graphicx}
\begin{document}
\title{Commissioning of the new taggers of the KLOE-2 
experiment
}
\author{
  {D.~Babusci, P.~Ciambrone, M.~Mascolo} \\{INFN-LNF, Frascati, Italy}\\
   ~\\
{R.~Messi} \\{Universit\`a e INFN Sezione di Roma ``Tor Vergata'' , Rome, Italy}\\
  ~\\
{D.~Moricciani} \\{INFN Sezione di Roma ``Tor Vergata'', Rome, Italy}\\
   ~\\
{S.~Fiore} \\{ENEA - Casaccia e INFN Sezione di Roma, Roma, Italy}\\
    ~\\
{P.~Gauzzi} \\{Universit\`a La Sapienza e INFN Sezione di Roma, Roma,
  Italy} \\
   ~\\
on behalf of the KLOE-2 Collaboration
}
\maketitle
\begin{abstract}
In order to fully reconstruct the $\gamma\gamma$ processes ($e^+e^- \to e^+e^- \gamma^{\star}\gamma^{\star}$) in the 
energy region of the $\phi$ meson production, new detectors along the DA$\Phi$NE 
beam line have been installed in order to detect the scattered $e^+e^-$.
\end{abstract}
  
\section{Introduction}
The renewed interest in $\gamma\gamma$ processes ($e^+e^- \to e^+e^-
\gamma^{\star}\gamma^{\star}\to e^+ e^- X$) is due to the possibility to
contribute to the calculation of the hadronic Light-by-Light (LbL) scattering
diagram that plays a relavant role in the theoretical evaluation of the
muon anomaly, (g-2)$_{\mu}$.
Accurate measurements of the radiative width of the pseudoscalar mesons
($X =\pi^0$ or $\eta$) and of the transition form factors at space-like $q^2$
can help in constraining models to be used in the LbL
calculation\cite{Jegerlehner:2009ry}.
Also the dipion final state ($X=\pi\pi$) can provide useful information,
according to recent dispersive approaches\cite{Colangelo:2014pva}, to the LbL calculation, as well
as to the study of the lightest scalar meson $f_0(500)$.\\
The trajectories of the off-momentum electrons from $\gamma\gamma$ events
have been studied by means of a Monte Carlo (MC) simulation based on BDSIM\cite{Agapov:2009zz}, to evaluate  
the exit point of the scattered particles from the DA$\Phi$NE beam pipe and to find proper location for the tagger 
devices. 
The results clearly indicate the need to place two different detectors in different regions 
on both sides of the interaction point (IP): the Low Energy Tagger (LET) to
detect leptons with energy  between 150 and 400 MeV and the High Energy
Tagger (HET) for those with energy greater than 420 MeV. 

\section{HET Detector}
\begin{figure}[h] 
	\begin{center}                               
		\includegraphics[width=.7\textwidth]{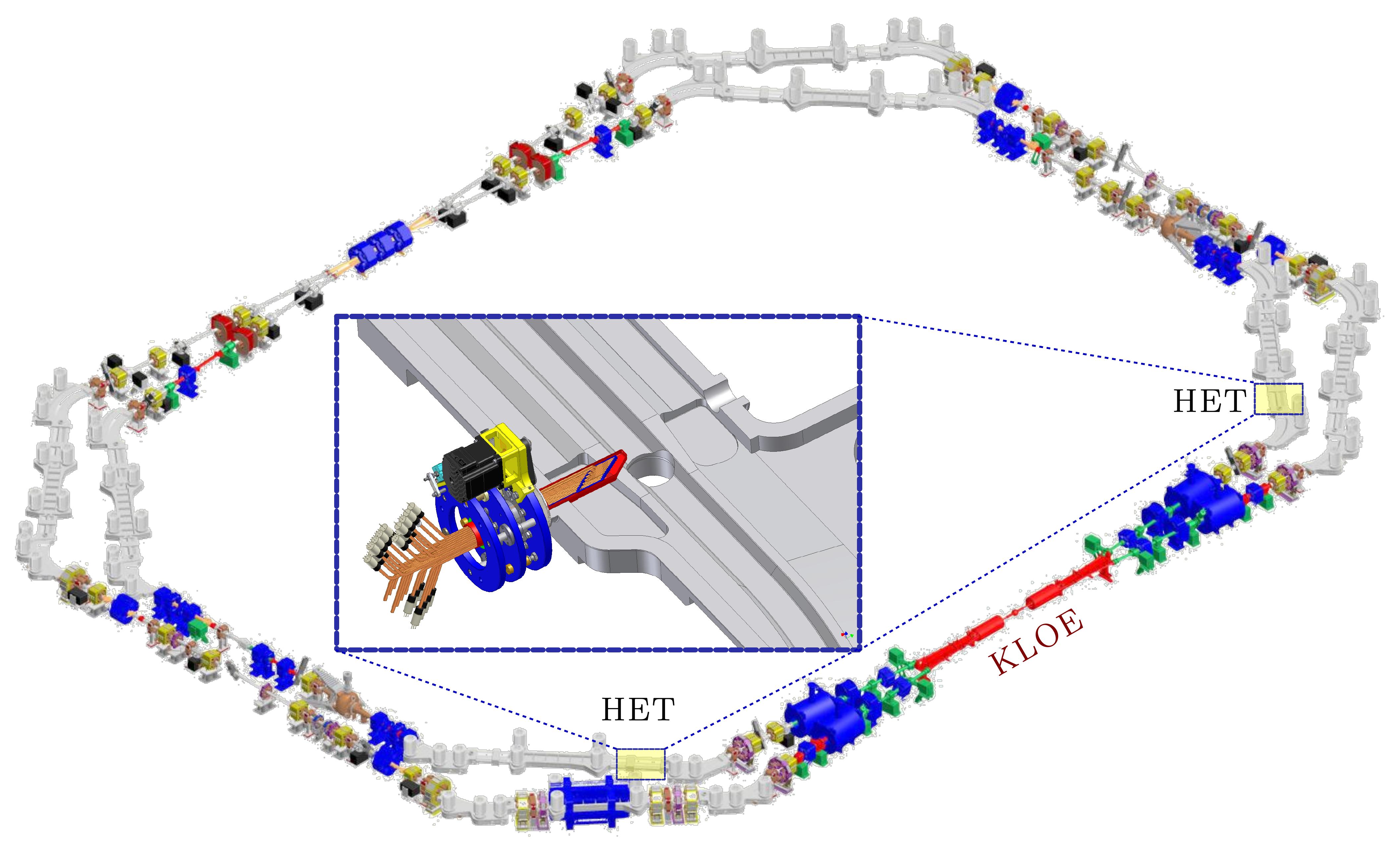}                
		\caption[Location of the HET detectors on DA$\Phi$NE]{Drawing of the two HET detectors placed on DA$\Phi$NE lattice.}
			\label{fig:taggers}                          
	\end{center}                                 
\end{figure}

The HET is a position detector used for measuring the deviation of
scattered $e^{\pm}$ from  their main orbit in DA$\Phi$NE. By means of 
this measurement and of its timing we are able to tag $\gamma\gamma$ events\cite{Babusci:2011bg}.
The two HET detectors are placed at the exit of the 
dipole magnets (see Fig. \ref{fig:taggers}), {11} {m} away from the 
IP, both on the positron and electron arms.
The sensitive area of the HET detector is made up of a set of 28 
plastic scintillators. The dimensions of each of them are $(\rm{3} 
\times \rm{5} \times \rm{6})~mm^3$.
One additional scintillator, of dimensions $(\rm{3} \times \rm{50} 
\times \rm{6})~mm^3$, is used for coincidence purposes.
The light emitted by each of the 28 scintillators is read out 
through plastic light guides by photomultipliers. The 28 scintillators 
are placed at different distances from the beam-line, in such a way that 
the measurement of the distance, between the hitting particle and the beam, 
can be performed simply knowing which scintillator has been fired.
They show their $(\rm{5} \times \rm{6})~mm^2$ face to the impinging particles 
that go through them along the thickness of \rm{3 mm}. The scintillators are 
not placed side by side, on the contrary there is an overlap of \rm{0.5 mm} on 
the \rm{5 mm} side.
The plastic scintillator used is the EJ-228 premium produced by Eljen Technology. 
Because of the small dimensions of the scintillator in use, the total 
light yield, due to a crossing electron or positron, is quite small. For this reason, we 
choose high quantum efficiency photomultipliers to minimize the 
probability of a particle to go undetected.
The photomultipliers used are compact size and high quantum efficiency ones,
model R9880U-110 SEL produced by Hamamatsu Photonics. 
The quantum efficiency is about 35\% for a wavelength going in the range 
from \rm{300~nm} to \rm{400~nm}, well matching the EJ-228 emission. 

\begin{figure}[h] 
\centering
		\includegraphics[width=0.7\textwidth, trim= 0 20mm 0 20mm, clip]{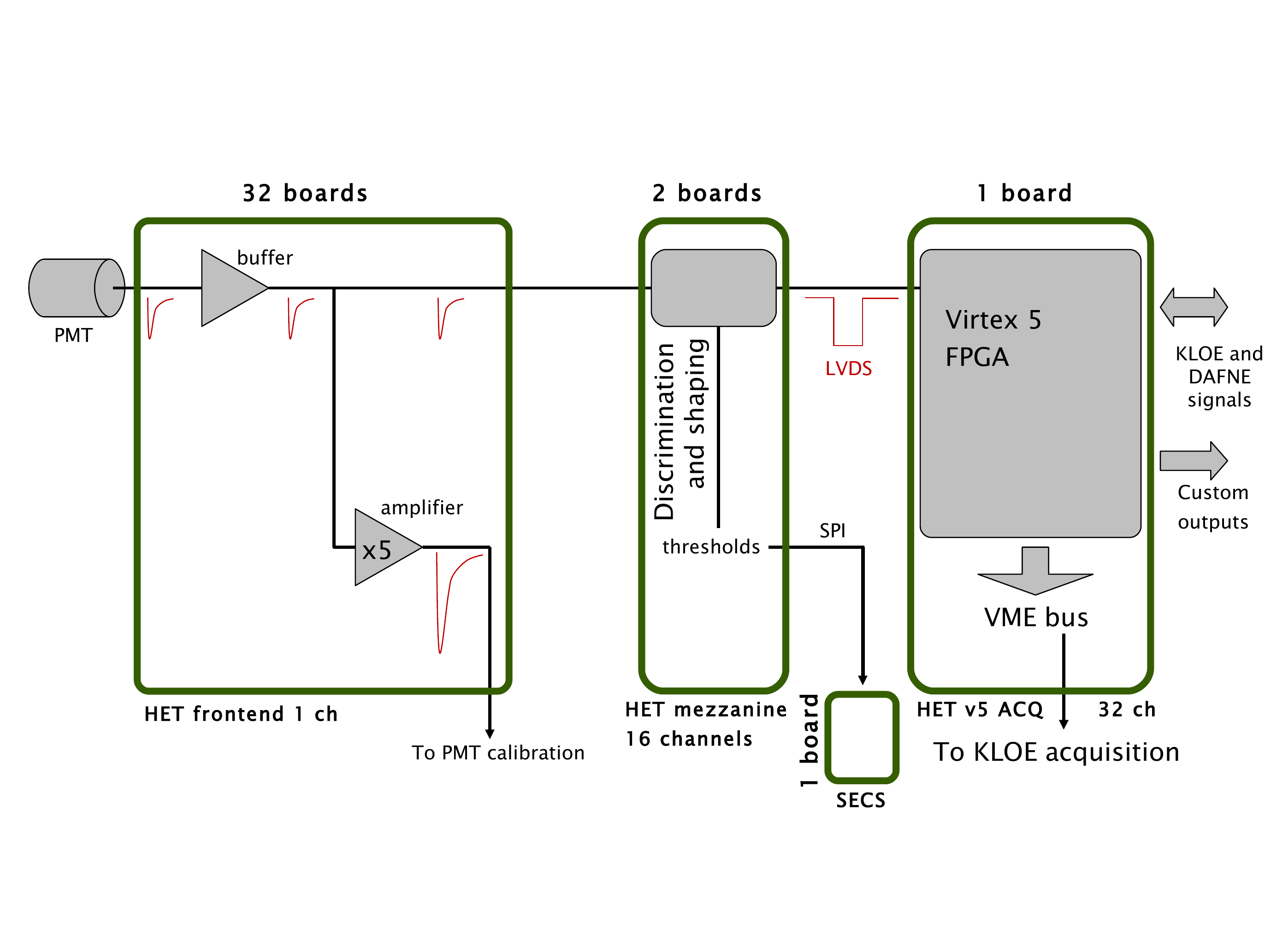}                
		\caption[Scheme of the HET electronic acquisition chain]{Scheme of the electronic 
		aquisition chain of the HET detector.}
		\label{fig:ELscheme}                          	
\end{figure}

The HET acquisition system is composed of a set of three electronic boards, and 
one slow control board (see the scheme in Fig. \ref{fig:ELscheme}). The first part 
of the chain, handling PMT analogue signals, is called the front-end electronics (FEE)
and is composed of the HET front-end board and the discrimination and shaping 
board. 
The HET main acquisition is a VME 6U board. The tasks handled by this board are 
to measure the timing of the signals coming from discriminators with respect to DA$\Phi$NE 
fiducial signal, store them only if a trigger from the KLOE main detector is asserted, and transmit data to the KLOE 
acquisition system through the VME bus. 
We developed a general purpose VME board that hosts a FX70T Virtex-5 FPGA
with 32 differential inputs for the TDC, an embedded DDR2 RAM  
memory for long data storage and many interfaces (VME, Ethernet, USB, RS232, Optical links) for 
readout, monitoring and debug. Thanks to the few resources used by the 4 Oversampling technique 
and since we did not have any limitation on the choice of the device, we were able to implement a 
complete DAQ system on the Virtex-5 FPGA comprehensive of a 32 TDC channels.
The TDC performs measurements continuously, producing a big amount of data but only a small 
fraction contains valid information. In order to select and store only relevant data, the KLOE trigger 
signals T1 and T2 are exploited and a zero suppression algorithm is
implemented to discard all the useless data.
\begin{figure}[htpb] 
\centering	
		\includegraphics[width=.7\textwidth]{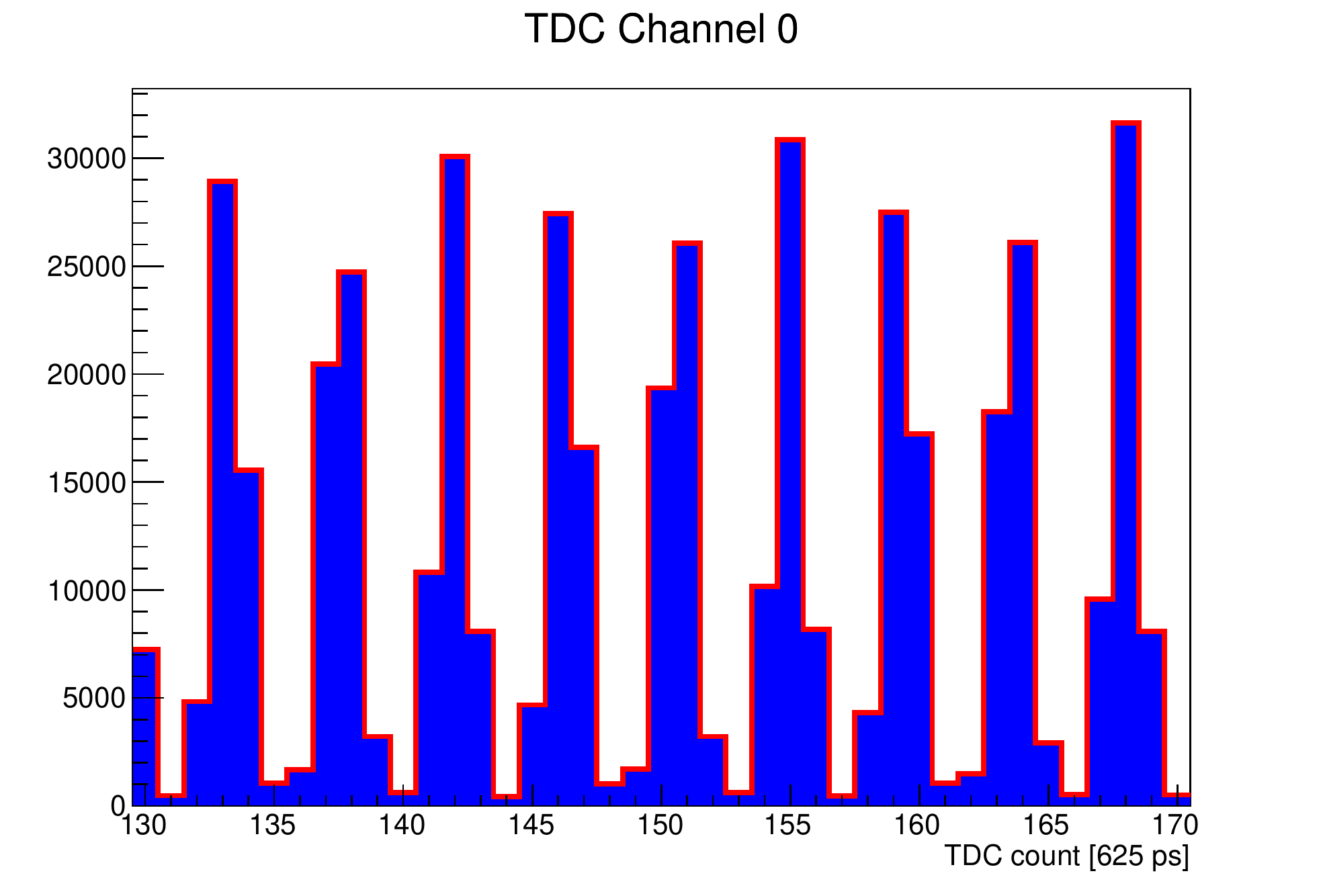} 
		\caption{TDC Spectrum for the long plastic scintillator.}             
		\label{fig:tdc}                          	
\end{figure}

Both electron and positron arm detectors are now installed and are in the commissioning phase.
Since the bunch crossing occurs in DA$\Phi$NE each $T_{bc} = $~2.7 ns, in order to properly 
disentangle leptons coming from two consecutive bunch crossings, the TDC time resolution 
must be less than $T_{bc}$ as it is shown in Fig.\ref{fig:tdc}.

\section{LET Detector}

\begin{figure}[htpb] 
\centering	
		\includegraphics[width=.7\textwidth]{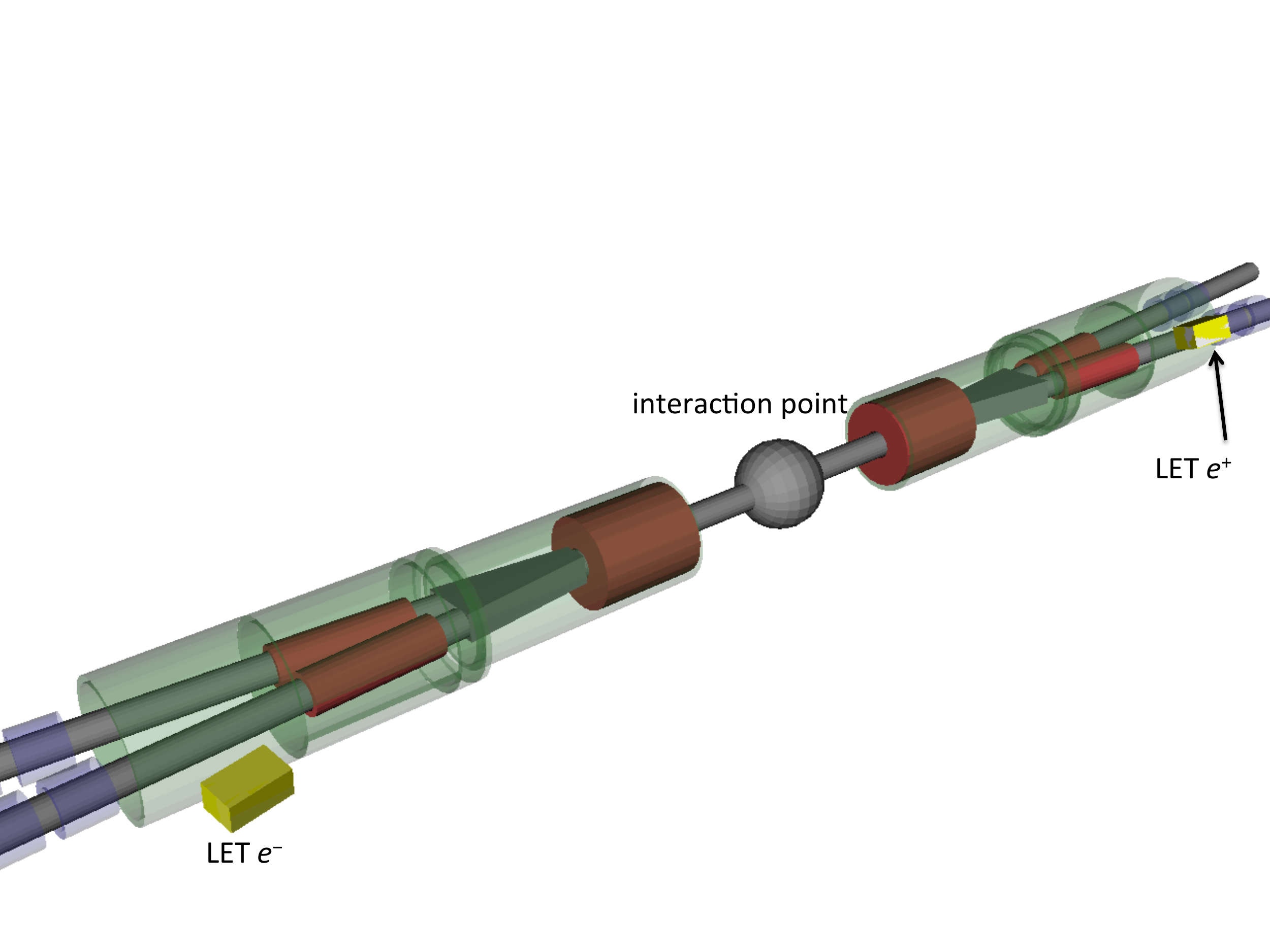} 
		\caption{Schematic view of the DA$\Phi$NE Interaction
                  region with the two LET calorimeters.}             
		\label{fig:letposition}                          	
\end{figure}

The LET consists of two devices simmetrically placed at about 1 m from the
DA$\Phi$NE IP (Fig.\ref{fig:letposition}).
The MC simulation showed that in that region there is only a rough
correlation between the energy and the trajectory of the scattered particle. 
For that reason a calorimetric detector has been choosen: each of the two
LET stations consists of an array of 5$\times$ 4 LYSO crystals, of
1.5$\times$ 1.5 cm$^2$ section and 12 cm length, pointing to the average
direction of the arriving particles, about 11$^o$ with respect to the beam
line.
The two stations are rotated by an angle of $\pm$ 17$^o$ with respect to
the horizontal plane, to maximize the number of collected positrons and
electrons, respectively. 
The choice of the optimal position has been performed with the help of a
MC simulation based on GEANT4 for the detector response and on BDSIM for
the particle tracking.\\
Each LYSO crystal is readout by a SiPM (S10362-33-025C by Hamamatsu,
3 $\times$ 3 mm$^2$, 14400 pixel).   
The FEE has been designed to be compatible with the KLOE electromagnetic
calorimeter (EMC) readout chain.
Its main features are: a very stable, low noise power supply for the SiPMs,
the possibility to set a working voltage for each channel in the range 60
-- 80 V, with 5 mV precision, and a low noise preamplifier.
Signals from preamplifier are sent to the KLOE EMC SDS (Splitter
Discriminator Shaper) boards, and then to ADCs and TDCs for charge and
arrival time measurements.\\ 
\begin{figure}[htpb] 
\centering	
		\includegraphics[width=.7\textwidth]{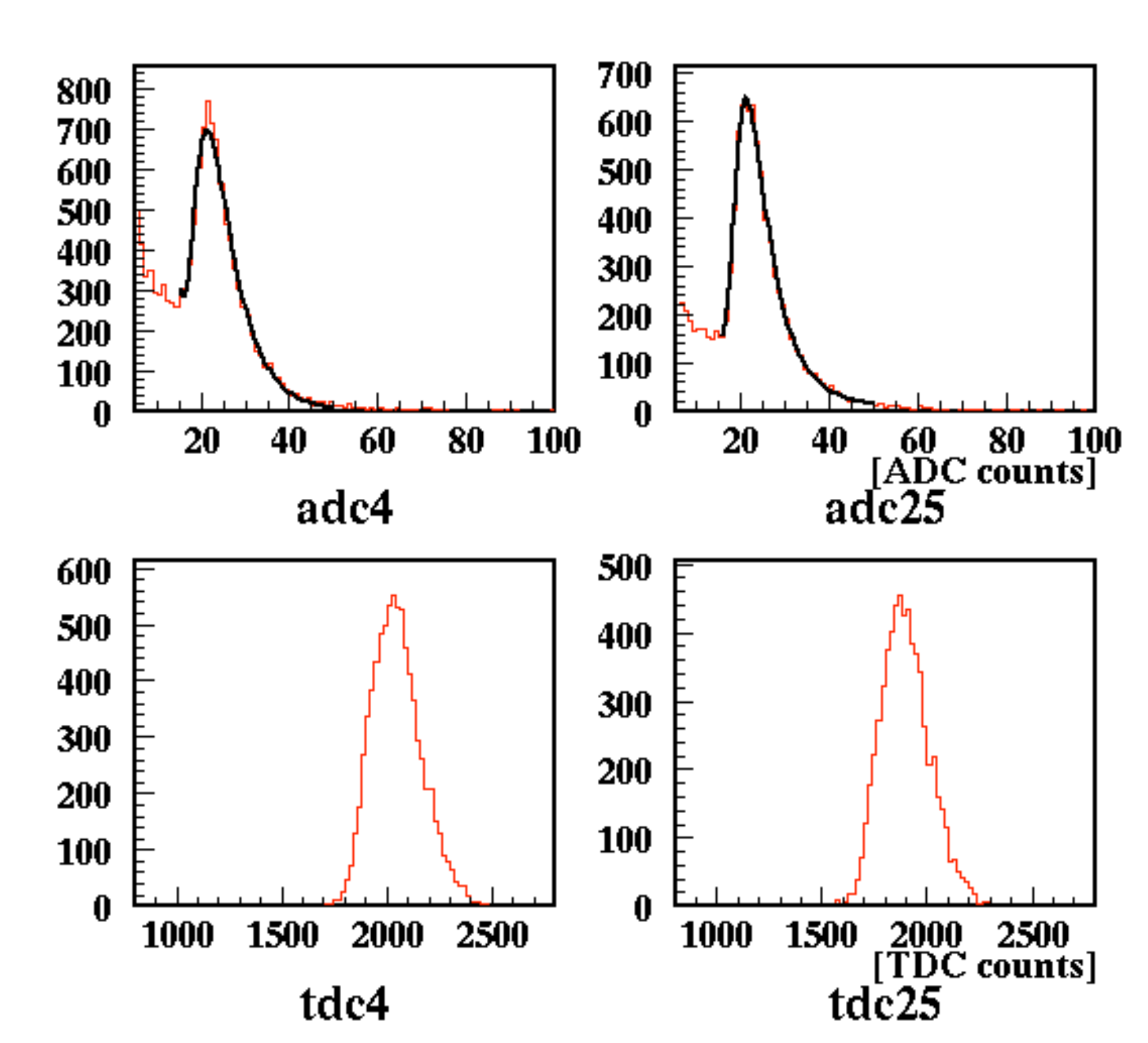} 
		\caption{Examples of ADC and TDC distributions for MIPs.}             
		\label{fig:letadctdc}                          	
\end{figure}
 The energy resolution of the LET calorimeter has been measured on a test
with electrons of energy between 50 and 500 MeV at the Frascati Beam
Test Facility, and turns out to be:
\begin{equation}
\frac{\sigma_E}{E}=\frac{2.4\%}{\sqrt{E[GeV]}}\oplus
6.5\%\oplus\frac{0.5\%}{E[GeV]}
\end{equation}
where the stocastic term corresponds to a collection of about 2
photoelectrons/MeV by the LYSO + SiPM system (no optical grease has been
used in the coupling).
The costant term was dominated by the leakage, because only the central part
of the calorimeter was readout in this test, and the third therm is due to
electronic noise.
This energy resolution well matches the requirement to be less than 10\%
in the range 150 -- 400 MeV\cite{Babusci:2009sg}.\\
The equalization of the response of the LET crystals and the timing
calibration is performed with minimum ionizing particles (MIPs) selected
by looking for high momentum tracks in cosmic rays collected without
circulating beams in DA$\Phi$NE.
Some examples of ADC and TDC distributions for MIPS are shown in
Fig.\ref{fig:letadctdc}.
The absolute energy scale calibration will be performed with radiative
Bhabha scattering ($e^+e^-\to e^+e^-\gamma$), with the photon and one
electron or positron reconstructed in the KLOE main detector, and the other
one detected in the LET.\\
A LED pulsing system has been installed to monitor relative SiPM gain
variations, and also temperature sensors are present on the two calorimeters.\\

\bibliography{taggers}
\end{document}